\documentstyle[12pt,epsfig,subfigure,amssymb]{article}
\begin{document}
\thispagestyle{empty}

\setcounter{page}{0}
\begin{flushright}
{CERN--TH/95--135\\
hep-ph/9506280}
\end{flushright}
\vskip1.5cm
\begin{center}

{\large{\bf THE PROTON SPIN PUZZLE AND\\
\vskip0.2cm
$\Lambda$ POLARIZATION IN DEEP--INELASTIC SCATTERING}}

\vskip1.5cm

{\bf John Ellis$^{a,}$\footnote{E-mail: johne@cernvm.cern.ch},
Dmitri Kharzeev$^{a,b,}$\footnote{E-mail: kharzeev@vxcern.cern.ch}
and Aram Kotzinian$^{c,d,}$\footnote{E-mail: aram@cernvm.cern.ch}}
\vskip0.8cm

{\it $ ^a)$ Theory Division\\
CERN\\
Geneva, Switzerland}
\medskip

{\it $^b)$ Physics Department\\
University of Bielefeld\\
33615 Bielefeld, Germany}
\medskip

{\it $^c)$ Yerevan Physics Institute\\
375036 Yerevan, Armenia}
\medskip

{\it $^d)$ INFN, Sezione di Trieste\\
Trieste, Italy}
\end{center}
\vspace{0.8cm}
\normalsize
\begin{abstract}
\normalsize
We point out that measurements of longitudinal
$\Lambda$ polarization in the target fragmentation region of
deep--inelastic $\nu \,N$ and $\mu \,N$ or $e \, N$ scattering may
test dynamical mechanisms invoked to explain the proton spin puzzle.
A previously-proposed model for polarized $\bar s s$ pairs in the
proton wave function reproduces successfully the negative $\Lambda$
polarization found in the WA59 $\bar \nu \, N$ experiment, and makes
predictions that could be tested in future $\mu \, N$ and $e \, N$
experiments.
\end{abstract}

\vspace {-0.10in}

\vskip2cm

\vspace{0.2cm}
CERN--TH/95--135

June 1995

\newpage
\section{Introduction}

$\;\;\quad$  Polarization measurements provide sensitive tests of models
of strong--interaction dynamics, and have produced a number of surprises.
The largest amount of discussion has been stimulated by measurements
of polarized deep--inelastic lepton--nucleon scattering structure
functions, which indicate that the angular momentum of the proton
is not distributed among its parton constituents in the way expected
in the na{\"{\i}}ve quark models. This discussion has been facilitated
by the relatively well--understood theoretical framework of perturbative
QCD, the operator product expansion and local operator matrix elements.
Other puzzling  polarization measurements, such as those of
helicity--amplitude effects in $pp$ elastic scattering or of
$\Lambda$ polarization in hadron--hadron and deep--inelastic scattering,
have suffered from the lack of such a clear basis for conceptual
analysis. Some modelling of nonperturbative
QCD is essential for the interpretation of these experiments, as well
as for understanding the local operator matrix elements determined by
deep--inelastic structure--function measurements.
\vskip0.3cm

Broadly speaking, one can distinguish two principal trends in the
interpretation of these deep--inelastic measurements. One notes that the
measured values of axial current matrix elements $\Delta q\
\cdot 2s_{\mu} = \langle p|\bar{q}\gamma_{\mu}\gamma_5 q|p\rangle$,
in particular the smallness of the singlet axial current
matrix element $\langle p|A^0|p\rangle = \Delta \Sigma \cdot 2 s_{\mu} :\
\Delta \Sigma = \Delta u + \Delta d + \Delta s$, are closer to the values
expected in chiral soliton models in the twin limits of massless
quarks and a large number of colours $N_c$, and interprets the data
in terms of the topology of the flavour $SU(3)$ group as reflected
in such chiral soliton models \cite{bek}.
The other trend assigns responsibility for the smallness of
$\langle p|A^0_{\mu}|p\rangle$ to the $U(1)$ axial--current
anomaly, which may a play either perturbatively, via a gluonic correction
to the polarized quark--parton distributions in the proton wave function:
$\Delta q \to \Delta q - (\alpha_s / 2\pi ) \Delta G$, or
nonperturbatively via a dynamical suppression of the $U(1)$ topological
susceptibility \cite{aec}.
\vskip0.3cm

The perturbative approach to suppression of the axial $U(1)$ matrix
element invokes a large polarized gluon distribution $\Delta G$
which is absent from the chiral soliton approach. These models are,
therefore, easy to distinguish in principle, and various proposals have
been made for measuring the gluon polarization via hard processes in
deep--inelastic scattering and elsewhere. Some experimental information is
starting to become available, but no test discriminating clearly between
these models has yet been made. At our present level of understanding,
any such test which goes beyond the perturbatively--calculable
hard--scattering framework necessarily involves additional nonperturbative
assumptions or inputs, for example in modelling the polarized proton
wave function.
\vskip0.3cm

One such model has recently been proposed \cite{ekks,aek},
in which a valence quark core
with (essentially) the na\"{\i}ve quark model spin content may be
accompanied by a spin--triplet $\bar{s}s$ pair in which the $\bar{s}$
antiquark is supposed to be negatively polarized, motivated by chiral
dynamics, and likewise the $s$ quark, motivated by $^3P_0$
quark condensation in the vacuum: $\langle 0|\bar{u}u, \bar{d}d,
\bar{s}s|0\rangle \neq 0.$ This model has been shown to
reproduce qualitatively experimental features
of $\phi$ production in $\bar{p}N$ annihilation and used to make further
predictions for $\phi$ and $f_2'$ production \cite{ekks},
and for depolarization in $\bar{p}p\to \bar{\Lambda}\Lambda$ \cite{aek}.
In the latter case, our model
predictions differ from those that could be expected in models with
a large polarized gluon distribution in the proton wave function.
\vskip0.3cm

In this paper we first point out that our polarized -- $\bar{s}s$ model
\cite{ekks,aek} provides a natural
interpretation of data on longitudinal $\Lambda$
polarization in the target fragmentation region in deep--inelastic
$\nu N$ collisions. In our model, the polarized $W$ emitted by the
$\nu$ selects a polarized quark from the target nucleon wave function.
Any $\bar{s}s$ pair in the remnant wave function would have the opposite
polarization, which can be transferred to final--state $\Lambda$
polarization.
This is indeed observed to be negative, and the polarization transfer
efficiency seems to be about $70\%$. A polarized--gluon model appears
likely to yield the opposite sign of $\Lambda$ polarization in the target
fragmentation region.
\vskip0.3cm

We then use our model to make predictions for $\Lambda$ polarization
in deep--inelastic $\mu N$ or $e \, N$ collisions. In this case,
the polarized $\mu$ or $e$ beam emits a virtual photon with non--zero
longitudinal polarization,
which in turn selects preferentially one polarization state of the struck
quark. In our model, the opposite polarization of a remnant $\bar{s}s$
pair can again be transferred to final--state $\Lambda$ polarization,
with the efficiency extracted from $\nu N$ collisions. We present
detailed predictions for $\Lambda$ polarization in the target
fragmentation region for both polarized and unpolarized targets, and also
make predictions for correlations between $\Lambda$ and $\bar{\Lambda}$
polarization measurements in the current and target fragmentation regions.
\vskip0.3cm

The layout of this paper is as follows. Our model for polarized $q$
distributions
in the nucleon wave function is reviewed in Section 2. Our interpretation
of the $\nu N$ data is presented in Section 3, accompanied by
quantitative results for different kinematic regions obtained using the
LEPTO Monte Carlo program.
This program is then used in Section 4 to make quantitative predictions
for the target and current fragmentation regions in $\mu \,N$ and
$e \, N$ scattering, and correlation measurements are discussed
in Section 5.
Finally, in Section 6 we draw some conclusions and compare
our predictions with what might be expected qualitatively in a
polarized--gluon model.
\vskip0.3cm

\section{Model for polarized quark distributions}

$\;\;\quad$ Results from the deep--inelastic scattering experiments
\cite{DIS} clearly indicate
that the contribution of light quarks to the spin of the proton is small:
$\Delta \Sigma = \Delta u + \Delta d + \Delta s \ll 1$.
This comes as a surprise from
the point of view of the na\"{\i}ve constituent quark model, which would
suggest $\Delta \Sigma \simeq \Delta u + \Delta d \simeq 1,$
and $\Delta s = 0$. The experimental data also allow to extract
the separate contributions of $u,d$ and $s$ quarks
\cite{EK}:
\begin{equation}
\Delta u = 0.83\pm0.03,\ \Delta d = - 0.43\pm0.03,\
\Delta s = - 0.10\pm0.03.
\label{Delta}
\end{equation}
Strange quarks therefore appear to have a net polarization opposite to
the proton
spin. Though the matrix elements $\Delta u$ and $\Delta d$ involve both
valence and sea quarks, the small value of $\Delta \Sigma$ suggests that
the light quark sea is also negatively polarized, so as to compensate
to large extent the spin carried by the valence quarks.
\vskip0.3cm

The origin of this negative sea polarization is likely nonperturbative
and still has to be understood. One model motivated by chiral dynamics
has recently been proposed \cite{ekks,aek}.
It is based on two major observations. First,
the fact that the masses of Goldstone bosons of spontaneously broken
$SU(3)_L \times SU(3)_R$ chiral symmetry -- pions and kaons -- are small
at the typical hadronic scale can
be attributed to the existence of effective strong attraction between
quarks and antiquarks in the
pseudoscalar $J^{PC}=0^{-+}$ channel. Secondly, from phenomenological
analyses of the quark condensates in the framework of the QCD sum rules
\cite{SVZ}, \cite{str} it is known
that the density of quark-antiquark pairs in nonperturbative vacuum
is quite high:
\begin{equation}
\langle 0|\bar{u}u|0\rangle\simeq
\langle 0|\bar{d}d|0\rangle\simeq (250 MeV)^3,\
\langle 0|\bar{s}s|0\rangle\simeq
(0.8\pm 0.1) \langle 0|\bar{q}q|0\rangle. \label{cond}
\end{equation}
It is worthwile to note that Eq(\ref{cond}) indicates that
the density of strange quarks in the vacuum is
comparable to the density of $u$ and $d$ quarks.
\vskip0.3cm

Let us now consider, following \cite{aek}, the basic $|uud\rangle$
proton state
immersed in the QCD vacuum. The strong attraction in the spin-singlet
pseudoscalar channel discussed above will induce correlations between
 valence quarks from the proton
wave function and vacuum antiquarks with opposite spins.
As a consequence of this, the spin of the antiquarks will be
aligned {\it opposite} to the proton spin. Moreover, we note that
in order to preserve the vacuum quantum numbers ($J^{PC}=0^{++}$),
quark-antiquark pairs must be in a relative spin-triplet,
$L=1$ $^3P_0$ state.
Therefore the spin of vacuum sea quarks must also be aligned
{\it opposite} to the proton spin.
The resulting wave function of the proton will therefore contain
negatively polarized $\bar{q}q$ components, which will effectively
decrease the fraction of the proton spin carried by quarks. The negative
contribution to the spin will be compensated by a positive contribution
from orbital angular momentum, so the total angular momentum sum rule
of course will be satisfied. The proposed mechanism can be interpreted
as an effective vacuum ``screening" of the quark (and proton) spin
in the nonperturbative vacuum with spontaneously broken chiral symmetry.
\vskip0.3cm

In view of (\ref{cond}), this mechanism should be applicable to
$\bar s s$ components as well as $\bar u u$ and $\bar d d$. Therefore
we expect that the proton wave function will
contains a considerable admixture with an $\bar{s}s$ component:
\begin{equation}
|p\rangle = v \sum^{\infty}_{X=0} |uudX\rangle +
z \sum^{\infty}_{X=0} |uud\bar{s}sX\rangle
+ \cdots,
\label{wave}
\end{equation}
where $X$ denotes Fock space components not containing $\bar s s$ pairs,
and the dots denote components with two or more $\bar s s$ pairs, which
we assume to be negligible so that $|v|^2+|z|^2 \simeq 1$.
The simplest wave function of the $\bar{s}s-$ containing
component, consistent with the dynamical chirality and spin arguments
 discussed above,
corresponds
to a spin-triplet, polarized $S_z=-1$ $\bar{s}s$ pair with
angular momentum $L_z=+1$ coupled to the ``usual"
$S_z=1/2$ $|uud\rangle$ valence state.
\vskip0.3cm

Within this general framework, it is possible to imagine that each
constituent quark, in a na\"{\i}ve quark model for the proton wave
function,contains a valence $u$ or $d$ quark with the same polarization
as the parent constituent quark, and its own individual sea of
$\bar q q$ pairs with the above-mentioned spin correlations. In a
deep-inelastic scattering event in which one of the valence quarks is
struck, it might be a good approximation to consider its parent
constituent quark as being the only one dissociated, with the other
constituent quarks left essentially intact as spectators. In this
refinement of the model of \cite{aek}, the polarization of the remnant
$\bar s s$ pair would be 100\% anticorrelated with that of the struck
valence quark (or its parent constituent quark).
\vskip0.3cm

The ``experimental" value (\ref{Delta}) suggests that
\begin{equation}
|z|^2 \simeq - \Delta s \simeq 0.10,
\end{equation}
which is consistent with the limits derived from the phenomenology of
$\bar{s}s$ production in $\bar{p}N$ annihilation \cite{ekks}:
\begin{equation}
0.01 \leq |z|^2 \leq 0.19. \label{adm}
\end{equation}
This polarized intrinsic strangeness model has been shown to reproduce
qualitatively the channel--dependent, non--universal excess
of $\phi$ production
in $\bar{p}N$ annihilation at rest observed recently at LEAR \cite{phi}.
Within this approach,
a large apparent violation of the OZI rule is interpreted in terms of
``rearrangement" and ``shake--out" of an intrinsic $\bar{s}s$ component
of the nucleon wave function.
\vskip0.3cm

The self--analyzing properties of $\Lambda$ ($\bar{\Lambda}$) make this
particle especially interesting for spin physics. This hyperon can be
used as a $s$ quark polarimeter since the polarization of $\Lambda$
is defined by polarization of its $s$ quark.
The production of $\bar{\Lambda}\Lambda$ pairs in $\bar{p}p$ annihilation
in our model is viewed \cite{ekks} as the dissociation of a spin-triplet
$\bar{s}s$ pair from the initial proton or antiproton into a
$\bar{\Lambda}\Lambda$ state.
Since the spin of the $\Lambda$ is carried by the spin of
the strange quark,  this (spin--correlation--conserving) dissociation
leads to a spin--triplet final state for the two hyperons. This is indeed
consistent with the experimental observation \cite{PS185} that the
spin--singlet fraction in the $\bar{\Lambda}\Lambda$ final state is
equal to zero within statistical errors.
 Since the initial $\bar{s}s$ pair
carries a polarization opposite to the (anti)proton spin, the model also
predicts
that the spin of the final $S=1$ $\bar{\Lambda}\Lambda$ pair is polarized
in the direction opposite to the spin of initial spin-triplet $\bar{p}p$
state, and therefore the depolarization $D_{nn}$ is negative.
This prediction can be tested experimentally in the near future
\cite{Kla}. It is, however, important to test the polarized intrinsic
strangeness model in the same kinematical conditions in which the
contribution of the strange quarks to the spin of the proton $\Delta s$
is measured, i.e., in deep--inelastic scattering. In the next
Sections we will discuss the consequences of this model for the
polarization of the $\Lambda$'s produced in deep--inelastic scattering.
\vskip0.3cm

\section{Polarization of $\Lambda$'s in the Target Fragmentation
Region in Deep-Inelastic $\bar \nu$ Scattering}
\vskip0.3cm

$\;\;\quad$  Measurements of $\Lambda$ polarization have been made
in the target fragmentation region ( $x_F<0$ ) in $\nu$ and $\bar \nu$
deep-inelastic scattering experiments~\cite{wa59}-\cite{ammosov}.
Non-trivial {\it negative} longitudinal polarization, measured with
respect to the direction of the momentum transfer from the beam,
has been observed in two experiments~\cite{wa59}-\cite{jones},whereas
the data on transverse $\Lambda$ polarization are quite contradictory.
The longitudinal-polarization data of the WA59 experiment~\cite{wa59}
have the  best statistical accuracy, and in this Section we present
an interpretation of these data using the simple model of the polarized
nucleon wave function described in Section 2.
\vskip0.3cm

The essence of our argument is that the right-handed polarization
of the $\bar \nu$ beam is transferred to the hadrons via polarized
$W^-$-exchange, which selects preferentially one {\it longitudinal}
polarization state of the nucleon target, which is reflected in
nontrivial {\it longitudinal} polarization of $\Lambda$'s produced
in the target fragmentation region. Specifically, in most interactions
the $\bar \nu$-induced $W$ removes a {\it positively}-polarized
$u$ quark from the nucleon target, as seen in Fig. 1.
The na\"{\i}ve model
of Section 2 then predicts that any $s$ quark in the target fragment
should have {\it negative} longitudinal polarization, so that the
longitudinal $\Lambda$ polarization should also be {\it negative},
as observed.

\begin{figure}[htb]
\begin{center}
\mbox{ \psfig{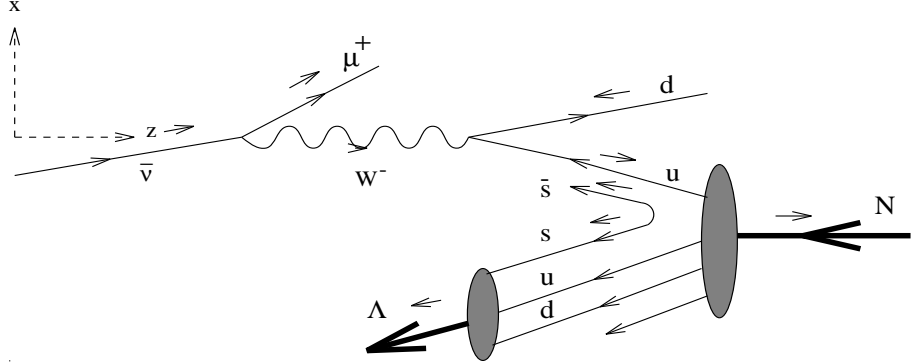} }
\end{center}
\caption{\it Dominant diagram for $\Lambda$ production in the
target fragmentation region due to scattering on a valence $u$ quark.
Each small arrow represent the longitudinal polarization
of the corresponding particle.}
\label{fig:uquark}
\end{figure}
\vskip0.3cm

To be more quantitative, we denote by $P_{s\pm q}$ ( $P_{s\pm \bar q}$ )
the probability that the longitudinal projection of the remnant $s$ quark
spin is parallel/antiparallel to that of the struck quark $q$
(antiquark $\bar q$), and define the spin-correlation coefficient

\begin{equation}
c_{s \,q}=\frac{P_{s +\,q} - P_{s -\,q}}{P_{s +\,q} + P_{s -\,q}}.
\label{cdef}
\end{equation}
\vskip0.3cm

In the na\"{\i}ve quark-parton model of deep-inelastic
$\nu$ or $\bar \nu$ scattering, the net longitudinal
polarization of a remnant $s$ quark , $P_{s}$, is given by
\begin{equation}
P_{s}=\frac{\sum_q c_{s\,q}N_q -
\sum_{\bar q} c_{s\,\bar q}N_{\bar q}}{N_{tot}},
\label{ps}
\end{equation}
where $N_q$ ($N_{\bar q}$) is the total number of events selected
in which a quark (antiquark) is struck, and $N_{tot}=N_q+N_{\bar q}$
is the total number of events selected. The antiquarks contribute with
a negative sign because their charged-current weak interactions are
righthanded. The relative proportion of $q$ and $\bar q$ events depend
in general on the range of the kinematic variables selected, leading to
dependencies of $P_s$ on the kinematic variables, as we discuss in more
detail below.
\vskip0.3cm

According to the simple polarization model discussed in the previous
section, the polarization of the remnant $s$ quark is 100\%
{\it anticorrelated} with that of the valence quark, and 100\%
{\it correlated} with that of struck sea $\bar s$ antiquark (see Fig. 2):

\begin{figure}[htb]
\begin{center}
\mbox{ \psfig{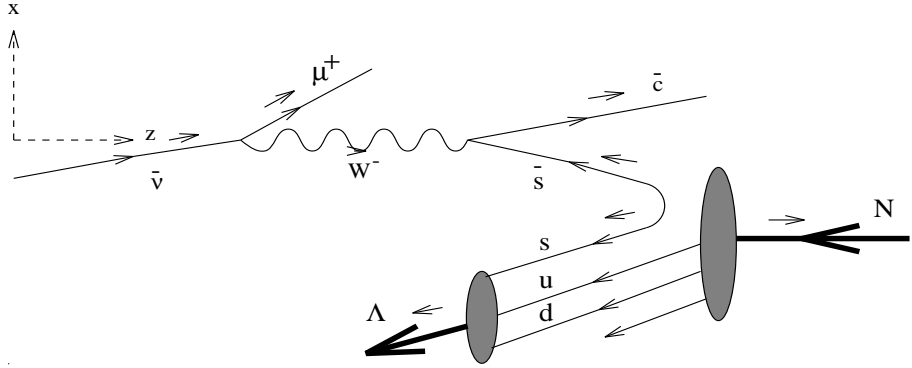} }
\end{center}
\caption{\it Diagram for $\Lambda$ production in a $\bar \nu \, N$
event due to W interaction with a $\bar s$ quark from the sea. As in
Fig. 1, the small arrows represent longitudinal polarizations. }
\label{fig:cquark}
\end{figure}

\begin{eqnarray}
\label{ccss}
c_{s\,u_{val}} &=& c_{s\,d_{val}}=-1,\nonumber\\
c_{s\,\bar q_{sea}} &=& \delta_{\bar s,\bar q}.\
\end{eqnarray}
The correlation of the remnant $s$ quark polarization with that of any
other struck sea quark depends whether they come from the same parent
constituent quark. If yes, which might be the dominant case, we would
expect

\begin{equation}
\label{csqs}
c_{s\, q_{sea}}=1.
\end{equation}
If no, however, the correlation would be reduced. The predictions we
present later are insensitive to the value of $c_{s\, q_{sea}}$ in the
region of Bjorken $x>0.15$, and also for lower $x$ in $\bar \nu \, N$
scattering. For definiteness we use (\ref{csqs}) in the following.
\vskip0.3cm

According to the simple quark model
of the polarized $\Lambda$ wave function,
the polarization of a directly-produced $\Lambda$ is the same as that of
the remnant $s$ quark. However, final-state $\Lambda$'s may also be
produced indirectly via the decays of heavier hyperon resonances,
which tends to dilute the $\Lambda$ polarization by a factor we denote
by $D_F$. Thus the final-state longitudinal $\Lambda$ polarization is
\begin{equation}
P_{\Lambda}=D_F P_{s},
\label{plam}
\end{equation}
where $P_s$ is given by equation (\ref{ps}). The fraction of $\Lambda$'s
produced indirectly may vary with the kinematical conditions, e.g., it
may be higher when the invariant mass of the produced hadron
system is larger.
\vskip0.3cm

We have used the latest version of the Lund Monte
Carlo program LEPTO6.2~\cite{ing}, with default values of parameters,
to obtain numerical results.
This program provides a good description of the existing data on
unpolarized
semi--inclusive hadron production in deep--inelastic scattering.
We implement into it the  spin correlations following from our model
(see (\ref{ccss}), (\ref{csqs})).
We have generated a sample of
deep-inelastic events with the $\bar \nu$ energy set equal
to the average value in the WA59 experiment, using
its target composition and kinematical
cuts, and then selected events with $\Lambda$'s produced in the
target fragmentation region. We present in Table 1 our results for the
polarization $P_s$ of the remnant $s$ quark in various ranges of
Bjorken $x$, together with the corresponding values of $P_{\Lambda}$
measured in WA59 experiment. We also tabulate the corresponding values
of $D_F$ inferred from our calculated values of $P_s$ and the measured
values of $P_{\Lambda}$.

\begin{table}[htb]
\begin{center}
\begin{tabular}{|c||c|c|c|} \hline
$x$ range      & $0 < x < 1$      & $0 < x < 0.2$    & $0.2 < x < 1$    \\
\hline \hline
$P_{\Lambda}$ in WA59 experiment
& $-0.63 \pm 0.13$ & $-0.46 \pm 0.19$ & $-0.85 \pm 0.19$ \\
\hline
$P_{s}$ in our model
& $-0.86$           & $-0.84$           & $-0.94$           \\
\hline
Dilution factor $D_F$
& $ 0.73 \pm 0.15$ & $ 0.55 \pm 0.23$ & $ 0.90 \pm 0.20$\\
\hline
\end{tabular}
\end{center}
\caption{\it $\Lambda$ polarization in the target fragmentation
region ($x_F < 0$).}
\label{tab:antineutrino}
\end{table}
\vskip0.3cm

The expected remnant $s$ quark polarization is smaller in the low-$x$
region ($x<0.2$) where the relative weight of events with struck
sea $u$, $\bar d$ and $c$ quarks is higher than in the valence-quark
region ($x>0.2$). Comparing the values of $P_{s}$ and
$P_{\Lambda}$ in this valence region, we
conclude that the polarization of $\bar s s$ pair must indeed be highly
correlated with that of the valence quark:
$c_{s\,u_{val}} \approx 1$ as we have assumed.
\vskip0.3cm

The lower value of $P_{\Lambda}$ in the sea region ( $X<0.2$ ) may be
due in part to the larger relative weight of struck sea-quark events,
and in part to a larger fraction of $\Lambda$'s being produced
indirectly via the decays of heavier hyperons in this $x$ range, where
the invariant hadronic mass is larger.
\vskip0.3cm

We are gratified that our na\"{\i}ve model
describes correctly the {\it sign}, {\it order of magnitude}
and {\it x dependence} of longitudinal $\Lambda$--polarization
in deep-inelastic $\bar \nu \, N$ collisions.
\vskip0.3cm

    It is interesting to contrast the above predictions of the
polarized $s \bar s$ sea model with what might be expected if
the EMC/SMC effect is essentially due to {\it positively}-polarized
gluons~\cite{aec}. In such a model, we would na\"{\i}vely
expect the depolarization
in $p \bar p \rightarrow \Lambda \bar \Lambda$ and the polarization of
$\Lambda$'s from target fragmentation the WA59 experiment (see Fig. 3)
to be {\it positive}, i.e., opposite in $sign$ to the
polarized $s \bar s$ sea model and to what was in fact observed
in the WA59 experiment. Detailed calculations in the
polarized gluon model lie, however, beyond the scope of this note.

\begin{figure}[htb]
\begin{center}
\mbox{ \psfig{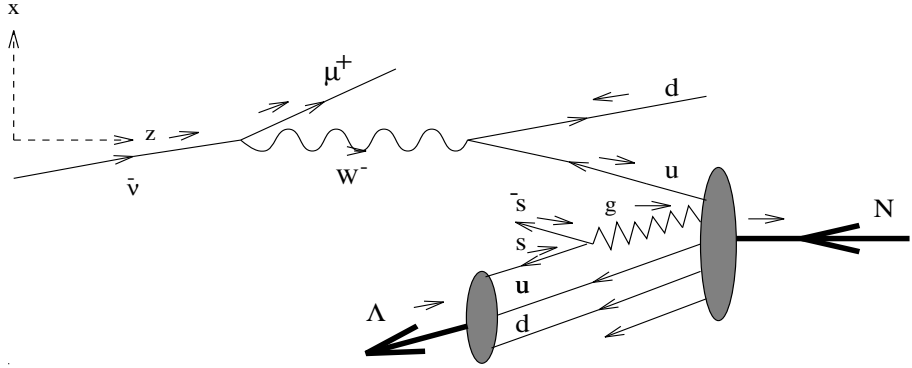} }
\end{center}
\caption{\it Possible $\Lambda$ polarization in deep-inelastic
$\bar \nu \, N$ scattering in the polarized gluon model.}
\label{fig:gluon}
\end{figure}

\section{Polarization of $\Lambda$'s in the Target Fragmentation Region
in Deep-Inelastic $\mu$ or $e$ Scattering}
\vspace{0.5cm}

$\;\;\quad$  Encouraged by the successful interpretation of WA59 data
in the previous Section, in this section we apply our model to predict
the polarization of $\Lambda$'s produced in the target fragmentation
region in the deep-inelastic scattering of polarized muons
on both unpolarized and polarized nucleon targets, as in the
experiment proposed at CERN \cite{hmc} and in polarized $e\,P$
scattering, as in the HERMES experiment \cite{herm}.
\vskip0.3cm

Our argument for $\mu \, N$ scattering
based on the fact that muon beams are naturally
longitudinally polarized, because the beam particles are produced
by charged-current weak decays. The degree of longitudinal
polarization, $P_B$, depends on the beam characteristics. Since
scattering via the electromagnetic interaction has different cross
sections for different quark longitudinal polarization states, the
struck quark (or antiquark) in the target has non-zero net
longitudinal polarization, and the model of the Section 2 then suggests
a {\it negatively-correlated} remnant $s$ quark polarization which
may be transferred to $\Lambda$'s produced in the target fragmentation
region. The observation by WA59 collaboration of large negative
longitudinal $\Lambda$ polarization in the target fragmentation region
of $\bar \nu\,N$ scattering suggests that the polarization transfer
mechanism does not involve strong dilution of the remnant $s$ quark
polarization.
\vskip0.3cm

We consider electromagnetic lepton-quark scattering in its
center-of-mass system.
with $z$ axis along lepton momentum. Depending on the relative sign
of the lepton and quark longitudinal spin projections, this process may
take place in the $s$-wave or in the $p$-wave.
In the $s$-wave case, the scattering
probability is independent of $y$, whereas in the $p$-wave case is
proportional to $(1-y)^2$, where $y$ is usual deep-inelastic energy
loss variable. The probability that the lepton  has
positive/negative longitudinal spin projection is
\begin{equation}
w_{\pm}^l=\frac{1}{2}(1 \pm P_B).
\label{wl}
\end{equation}
Analogously, the probability to find a quark (antiquark) with
nucleon-momentum fraction $x$ and
positive/negative longitudinal spin projection is
\begin{equation}
w_{\pm}^q(x) = \frac{1}{2}[q(x) \pm P_T \Delta q(x)],
\label{wq}
\end{equation}
where $P_T$ is the target polarization and $q(x)$ and $\Delta q(x)$ are
unpolarized and polarized quark distribution functions, respectively.
The lepton-quark interaction probability depends as follows
on the signs of the longitudinal spin projections
$w^{l\;\;q}_{i\;\;k} (i,k = +,-)$:
\newpage
\begin{eqnarray}
w^{l\;\;q}_{++} &=& e_q^2\; w^l_+ w^q_+(x) (1-y)^2 , \nonumber\\
w^{l\;\;q}_{+-} &=& e_q^2\; w^l_+ w^q_-(x)         , \nonumber\\
w^{l\;\;q}_{-+} &=& e_q^2\; w^l_- w^q_+(x) , \\
w^{l\;\;q}_{--} &=& e_q^2\; w^l_- w^q_-(x) (1-y)^2 , \nonumber\
\label{wlq}
\end{eqnarray}
where $e_q$ is the quark charge.

For the struck quark longitudinal polarization $P_q$, we get
\footnote{A more general expression for the struck quark polarization,
including effects of the intrinsic transverse momentum of quarks can be
found, for example, in \cite{ko}.}
\begin{equation}
P_q=\frac
{\sum_i(w^{l\;\;q}_{i\;+}-w^{l\;\;q}_{i\;-})}
{\sum_i(w^{l\;\;q}_{i\;+}+w^{l\;\;q}_{i\;-})}
=\frac{P_T\Delta q(x)-P_BD(y)q(x)}{q(x)-P_BP_TD(y)\Delta q(x)},
\label{pqq}
\end{equation}
where
\begin{equation}
D(y)=\frac{1-(1-y)^2}{1+(1-y)^2}
\label{dy}
\end{equation}
which is
commonly referred as longitudinal depolarization of virtual photon
with respect to parent lepton.
Using the spin correlation coefficient introduced in Section 3, it
is easy to find the following expression for
the polarization of the remnant $s$ quark
\begin{equation}
P_{s_{rem}}=\frac
{\sum_{i,q}[w^{l\;\;q}_{i\;+}-w^{l\;\;q}_{i\;-}]c_{s\,q}}
{\sum_{i,q}[w^{l\;\;q}_{i\;+}+w^{l\;\;q}_{i\;-}]}
=\frac {\sum_q\,e_q^2[P_T\Delta q(x)-P_BD(y)q(x)]c_{s\,q}}
{\sum_q\,e_q^2[q(x)-P_BP_TD(y)\Delta q(x)]},
\label{psr}
\end{equation}
where the summation over $q$ here means summation over
both quarks and antiquarks.
\vskip0.3cm

We see in formulae (\ref{pqq}) and (\ref{psr})
that there are two potential sources of
longitudinal polarization of the struck quark and remnant $s$ quark:
target polarization, and polarization
induced by interaction with the polarized beam. Note the nonlinear
dependence of expressions (\ref{pqq}) and (\ref{psr})
on beam and target polarization.
\vskip0.3cm

We have again used LEPTO6.2 to obtain numerical
results, by generating deep-inelastic scattering events with
$\Lambda$ production, using the beam energy and
kinematical cuts corresponding to those of the SMC
experiment. The remnant $s$ quark polarization was calculated according
to formula (\ref{psr}) with polarized quark distributions from the
paper by Brodsky, Burkardt and Schmidt~\cite{bbs}.
\vskip0.3cm

In the sea-quark region, the remnant $s$ quark polarization
is much smaller than in deep-inelastic $\bar \nu\,N$ scattering.
In the $\bar \nu$ case, events with a final--state
$\Lambda$ produced in the
target fragmentation region due to scattering on an $\bar s$
antiquark play an important role, and their contribution to $P_s$ has
the same sign as that from valence
$u$ quarks (see Fig. 1 and Fig. 2). In the $\mu$ case, the relative
weights of other sea quarks are higher, and contributions from valence
$u$ quarks and $\bar s$ antiquarks have opposite signs
(see Fig. 4 and Fig. 5).
Here, we present predictions only for the
valence-quark region: $x > 0.15$: for the reason mentioned above,
the calculations performed in the region $x<0.15$ would
yield smaller values of $\Lambda$ polarization, due to
a weaker spin correlation between the struck light sea quark and
the remnant $s$ quark, and would depend more on details of our model.
We also impose the $y$-cut $y>0.5$, so as to select events in which
the polarization transfer from the beam is big (see eqs.
(\ref{pqq}) and (\ref{dy})).

\begin{figure}[htb]
\begin{center}
\mbox{ \psfig{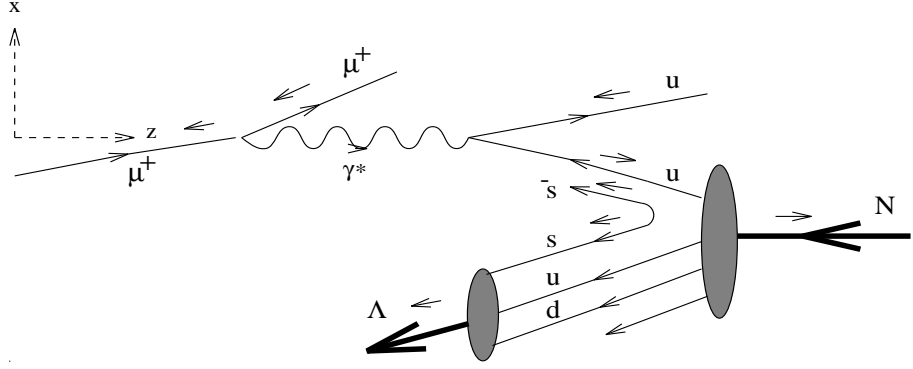} }
\end{center}
\caption{\it Dominant diagram for $\Lambda$ production in
deep-inelastic $\mu \, N$  scattering on a valence $u$ quark.}
\label{fig:umu}
\end{figure}

\begin{figure}[htb]
\begin{center}
\mbox{ \psfig{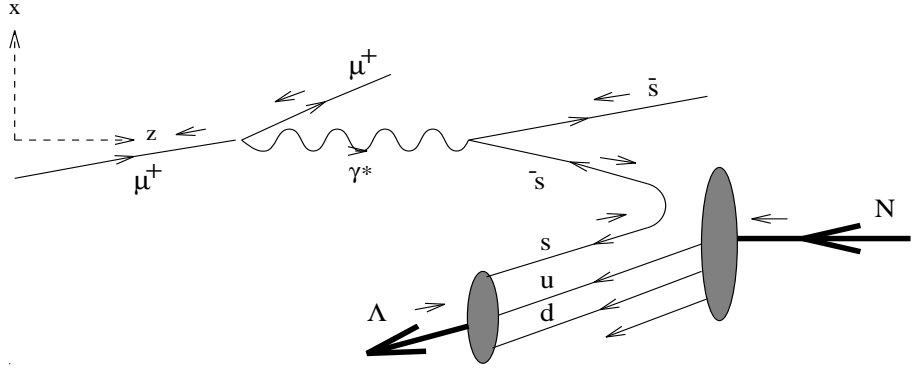} }
\end{center}
\caption{\it Diagram for $\Lambda$ production in a deep-inelastic
event due to $\mu$ interaction with an $\bar s$ antiquark from sea.}
\label{fig:cmu}
\end{figure}
\vskip0.3cm

For the new experiment proposed at CERN~\cite{hmc}, ammonia (NH$_3$)
is chosen as the target material.
We can treat this target approximately as an incoherent sum of
seven unpolarized neutrons, seven unpolarized protons and
three polarized protons (with polarization $P_T$).
The results of our calculations for a $\mu$ beam with the natural
longitudinal polarization $P_{\mu}=-0.8$ are shown in Fig. 6,
together with the cases $P_{\mu}=0$ and 0.8.
The spin correlation coefficients are taken from (\ref{ccss}).

\begin{figure}[htb]
\begin{center}
\mbox{\begin{tabular}[t]{cc}
\subfigure[{\it Proton target}]
{\psfig{file=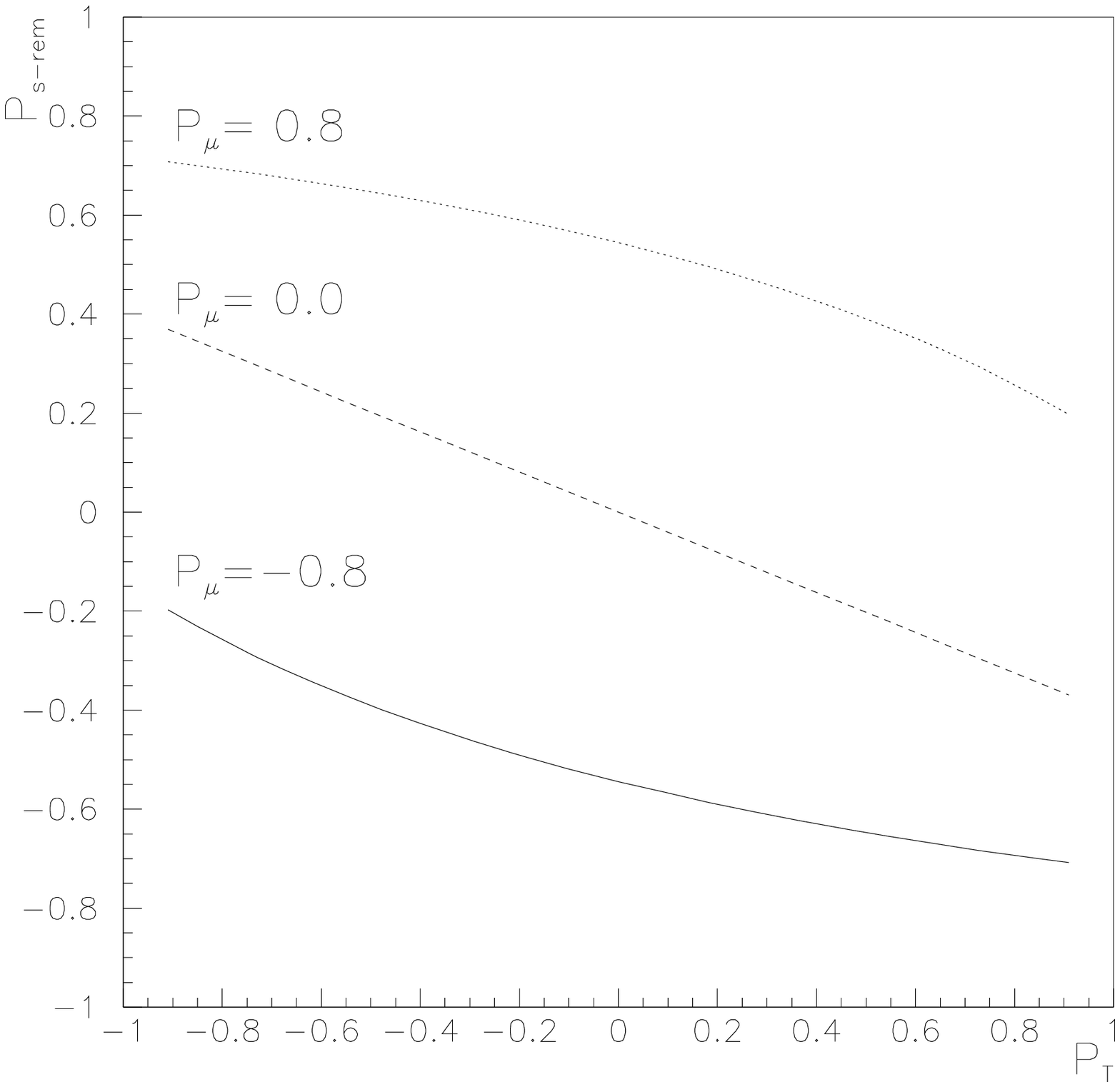,height=7cm,width=.45\textwidth}} &
\subfigure[{\it Ammonia target}]
{\psfig{file=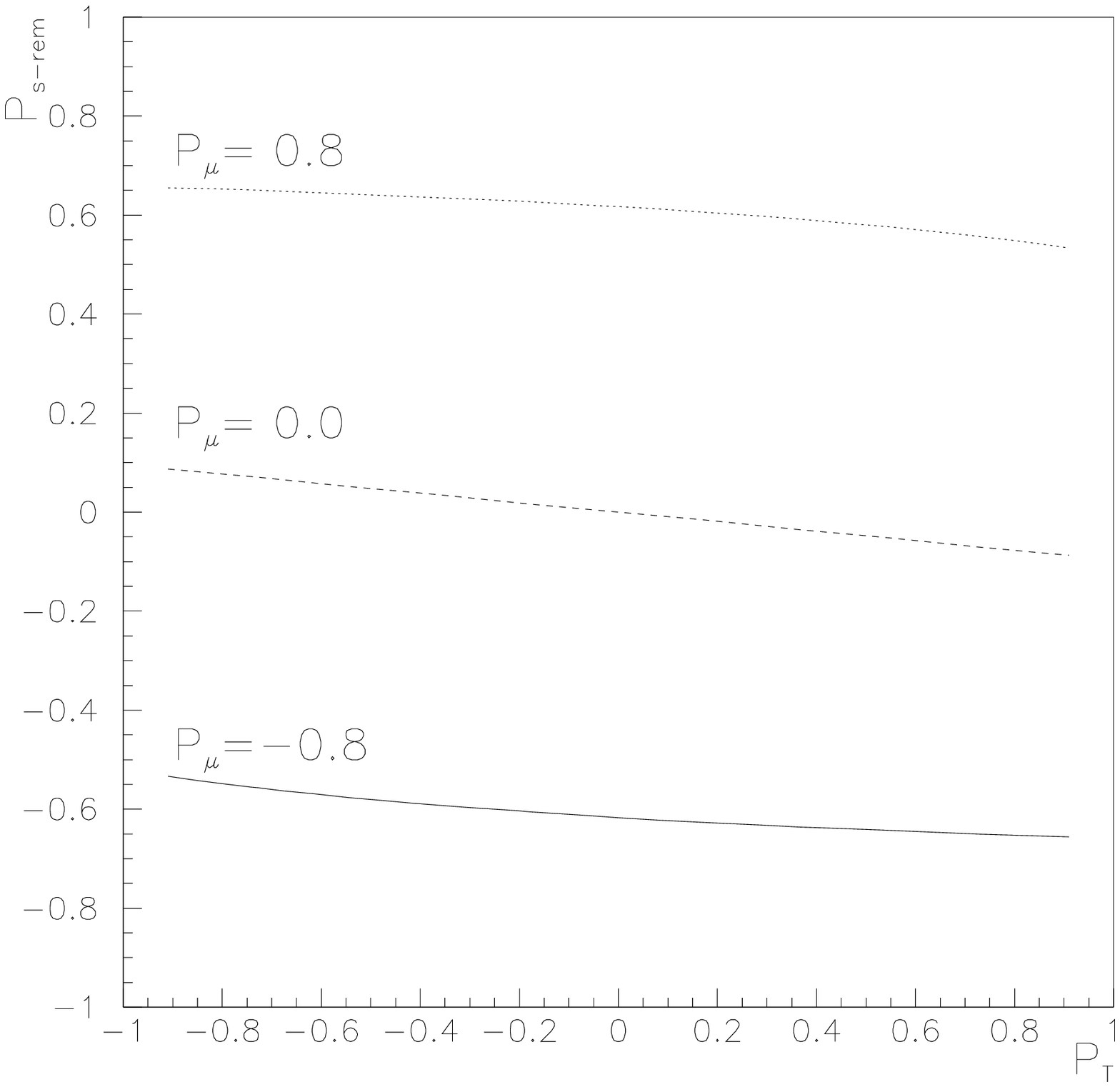,height=7cm,width=.45\textwidth}}
\end{tabular}}
\end{center}
\caption{\it Polarization of remnant s quark for deep-inelastic
$\mu$ scattering, as a function of the target polarization $P_T$ for
different values of the beam polarization $P_{\mu}$.
a) for a proton target, b)for an ammonia target. We assume $E_{\mu}$=190
GeV as in the proposed CERN experiment $[17]$ and the following cuts
were applied: $-0.3<x_F<0$, $x>0.15$, $0.5<y<0.9$.}
\label{fig:psmu}
\end{figure}
\vskip0.3cm
For deep-inelastic scattering on an ammonia target, most of the
remnant $s$ quark polarization $P_s$
is induced by polarization transfer from the lepton, and
the difference between $P_{s}$ for targets polarized with opposite
signs is smaller than in the proton target case.
\vskip0.3cm

The produced $\Lambda$ polarization is given by equation (\ref{plam}).
We concluded from our analysis of the WA59 data that
in the valence-quark region
the fragmentation dilution factor $D_F \gtrsim 0.7$. Therefore, we
expect large polarization effects also for $\Lambda$
production in the target fragmentation region in deep-inelastic
$\mu \,N$ scattering.
\vskip0.3cm

Calculations of the remnant $s$ quark polarization for $e \,P$
scattering, with $\Lambda$ produced in the target fragmentation region
and a beam energy $E_e=27.5$ GeV as in the HERMES experiment
\cite{herm}, are shown
in Fig. 7. The differences between Figs. 7 and 6 {\it a} are mainly
due to the different beam polarizations assumed. Energy-dependent
scaling violation effects are relatively unimportant.
\begin{figure}[htb]
\begin{center}
\mbox{ \psfig{file=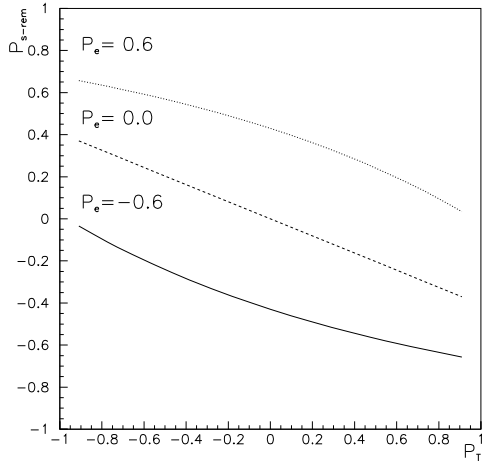,height=7cm,width=7cm} }
\end{center}
\caption{\it Polarization of remnant $s$ quark for deep-inelastic
polarized $e\, P$ scattering, as a function of the target
polarization $P_T$ for different values of the beam polarization.
A beam energy $E_e=27.5$ GeV as in the HERMES experiment $[18]$
was assumed, and the following cuts were applied:
$x_f<0$, $x>0.15$ and $0.5<y<0.85$.}
\label{fig:psherm}
\end{figure}

\vskip0.3cm

In the polarized-gluon model (see Fig. 8), we would na\"{\i}vely expect
the polarization of $\Lambda$'s to be {\it opposite} in sign from the
polarized $\bar s s$ sea model, though a detailed exploration goes
beyond the scope of this paper.

\begin{figure}[htb]
\begin{center}
\mbox{ \psfig{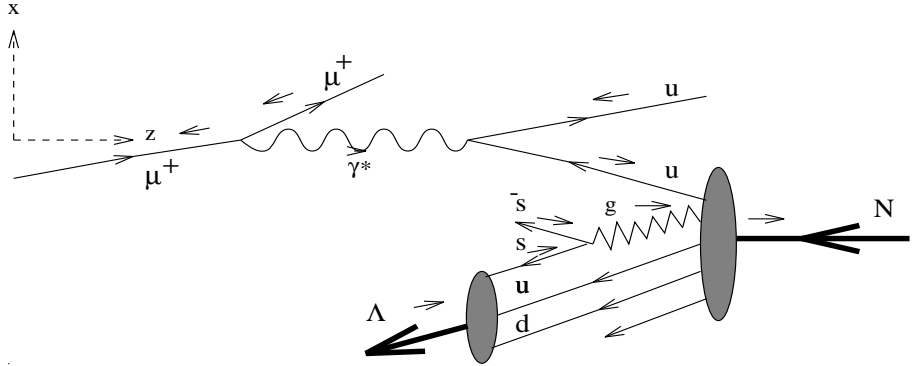} }
\end{center}
\caption{\it Possible $\Lambda$ polarization in the polarized-gluon
model.}
\label{fig:glmu}
\end{figure}

\section{ $\Lambda \, \bar \Lambda$ Polarization Correlation
Measurements}
\vspace {0.5cm}

$\;\;\quad$  It is interesting to consider also polarization effects in
the associated production of $\Lambda$ and $\bar \Lambda$, for which
we restrict our attention here to the unpolarized ammonia target case.
\vskip0.3cm

We first consider the case when a $\Lambda$ is produced in the target
fragmentation region and a $\bar \Lambda$ in the current
fragmentation region in $\mu \,N$ scattering. In
Table 4 we show the predicted $\Lambda$ and $\bar \Lambda$ polarizations
obtained using LEPTO6.2 Monte Carlo program with polarization effects
implemented as described previously.
Calculations are made assuming that there is no dilution ( $D_F=1$ )
in the polarization
transfers from the $s$ quark to the $\Lambda$ and from the
$\bar s$ antiquark to the $\bar \Lambda$.
We make the kinematical cuts $y>0.5$, $x_F^{\Lambda}<-x_F^0$
and $x_F^{\bar \Lambda}>x_F^0$ for various values of $x_F^0$.

\begin{table}[htb]
\begin{center}
\begin{tabular}{|c||c|c|} \hline
$\qquad x_F^0 \qquad $ & $ \qquad P_{\Lambda} \qquad $
&$ \qquad P_{\bar \Lambda} \qquad $ \\
\hline \hline
0.1 & -0.07 & -0.17\\
\hline
0.2 &  0.06 & -0.27\\
\hline
0.3 &  0.23 & -0.40\\
\hline
\end{tabular}
\end{center}
\caption{\it Polarization of associatively-produced
$\Lambda$ and $\bar \Lambda$ with $y>0.5$ and
$x_F^{\Lambda}<-x_F^0$ and $x_F^{\bar{\Lambda}}>x_F^0$,
for various choices of $x_F^0$.}
\label{tab:llb1}
\end{table}
\vskip0.3cm

As can be seen in Table 2, the maximal polarization values are found
when the $\Lambda$ and $\bar{\Lambda}$ are produced at high $x_F^0$.
In this case of large rapidity separation, the dominant mechanism
of associated production is shown in the diagram of Fig. 5,
with the $\bar{\Lambda}$ produced from the $\bar{s}$ antiquark
fragmentation. In the case of an unpolarized target,
the $\bar{s}s$ polarization state is determined by the spin transfer
from the polarized $\mu$. For this reason, the order of magnitude
and the signs of the polarizations of the
$\Lambda$ and $\bar{\Lambda}$ produced with  large
rapidity separations have to be the same also in the polarized-gluon
model.
\vskip0.3cm

Results for the case when both the $\Lambda$ and the $\bar \Lambda$
are produced in the target fragmentation region are shown in Table 3.
\begin{table}[htb]
\begin{center}
\begin{tabular}{|c||c|c|} \hline
$ \qquad x_F^0 \qquad $ & $ \qquad P_{\Lambda} \qquad $
& $ \qquad P_{\bar \Lambda} \qquad $ \\
\hline \hline
0.00 & -0.32 & -0.32\\
\hline
-0.10 & -0.44 & -0.44\\
\hline
-0.15 & -0.50 & -0.50\\
\hline
\end{tabular}
\end{center}
\caption{\it Polarization of associatively-produced
$\Lambda$ and $\bar \Lambda$ with $y>0.5$ and
$x_F^{\Lambda}<x_F^0$ and $x_F^{\bar \Lambda}<x_F^0$,
and for various choices of $x_F^0$}
\label{tab:llb2}
\end{table}
In this case, the dominant diagram for
$\Lambda$ -- $\bar \Lambda$ pair production is that of Fig. 4,
where the $\bar \Lambda$ produced in the target fragmentation region
inherits the polarization of the $\bar s$ quark.
At all values of $x_F^0$ the dominant mechanism of associated
$\Lambda -\bar{\Lambda}$ production is scattering on the valence
$u$ quark. For this reason, in contrast to the previous case,
in the polarized-gluon model we would expect the opposite sign for
the $\Lambda$ and $\bar \Lambda$ polarizations.

\section{Conclusions}
\vspace {0.5cm}
$\;\;\quad$  We have discussed in this paper ways in which measurements
of $\Lambda$ polarization in deep-inelastic final states may cast light
on the proton spin puzzle. Our suggestion is that $\Lambda$'s in the
target fragmentation region are likely to have longitudinal polarization
inherited from that of remnant $s$ quarks in the struck nucleon. We use a
simple model~\cite{ekks,aek} of the nucleon spin structure in which
the $s$ and $\bar s$ polarizations are anticorrelated with that of any
valence quark struck by the polarized lepton probe. Our model reproduces
successfully the sign, magnitude and $x$ dependence of the longitudinal
$\Lambda$ polarization measured in the WA59 deep-inelastic
$\bar \nu \, N$ experiment in the target fragmentation region.
We have also applied the model to make predictions for the longitudinal
polarization of $\Lambda$'s in the target fragmentation region in
deep-inelastic $\mu \, N$ scattering off either a polarized or an
unpolarized target, which are shown in Fig. 6, and polarized
deep-inelastic $e \,P$ scattering, as shown in Fig. 7.
\vskip0.3cm

A preliminary Monte-Carlo study shows~\cite{hmc} that the
proposed new CERN experiment could collect in one year of data taking
about $3\cdot 10^5$ events with $\Lambda$'s produced in the target
fragmentation region ( $-0.3<x_F<0$ ). After kinematical cuts $x>0.15$
and $y>0.5$, several thousand deep-inelastic events will be retained,
which is higher by an order of magnitude than in the WA59 experiment.
Taking into account the predicted degree of polarization, it seems that
the proposed new experiment at CERN~\cite{hmc} is capable of performing
measurements of $\Lambda$ polarization with sufficient accuracy
to test our model.
\vskip0.3cm

We emphasize that our model for the nucleon spin structure is not
rigorous, and represents an extrapolation of our present knowledge
of polarized structure functions~\cite{DIS}. It relies on the
interpretation of these data as due to polarized $\bar s s$ pairs in
nucleon~\cite{bek}, rather than polarized gluons~\cite{aec}. The
development of this latter model to make predictions for $\Lambda$
polarization in deep-inelastic scattering lies beyond the scope of
this paper. However, as we have indicated in previous Sections, it
seems to us likely that a plausible extrapolation of the
polarized-gluon model would predict the opposite sign for longitudinal
$\Lambda$ polarization in the target fragmentation region in
deep-inelastic $\bar \nu \,N$ and $\mu \, N$ or $ e \, N$
scattering. It seems to us
that it would be interesting for advocates of the polarized-gluon
interpretation of the polarized structure-function data to examine this
question. As we discussed in the previous paragraph, the statistics
for $\Lambda$ production in the proposed new experiment at
CERN~\cite{hmc} are likely to be sufficient to determine the sign of
the $\Lambda$ longitudinal polarization, and may therefore be able
to cast some light on the proton spin puzzle.

\section{Acknowledgements}
\vspace {0.5cm}
$\;\;\quad$  We thank Mary Alberg, Marek Karliner and Mikhail Sapozhnikov
for useful discussions. D.K. acknowledges the financial support of the
German Research Ministry (BMFT) under contract 06 BI 721.

\end{document}